\documentclass[aps,prl,twocolumn,groupedaddress]{revtex4}
\usepackage{epsfig}
\usepackage{pdfsync}

\def\q{{\bf q}}
\def\p{{\bf p}}

\def\8{\infty}
\def\oh{\frac{1}{2}}

\def\d{\partial}

\def\undertext#1{\vtop{\hbox{#1}\kern 1pt \hrule}}

\def\dbyd#1#2{\frac{d#1}{d#2}}

\def\be{\begin{equation}}
\def\ee{\end{equation}}
\def\bea{\begin{eqnarray} & &}
\def\eea{\end{eqnarray}}

\def\rf#1{(\ref{#1})}

\def\rf#1{(\ref{#1})}

\def\rfs#1{Eq.~\rf{#1}}

\begin{document}

%\preprint{}

\title{Feshbach molecule production in fermionic atomic gases}



%\author{Alexander Altland}
%\affiliation{Institut f\"ur Theoretische Physik,
%Universit\"at zu K\"oln, Z\"ulpicher Str 77, 50937 K\"oln, Germany}

\author{V. Gurarie}

%\thanks{}
%\affiliation{Kavli Institute for Theoretical Physics, University of California, Santa Barbara CA 93106 USA}
\affiliation{Department of Physics, University of Colorado,
Boulder CO 80309 USA}

\date{\today}

\begin{abstract}
This paper examines the problem of molecule production in an atomic fermionic gas close to an $s$-wave Feshbach resonance by means of a magnetic
field sweep through the resonance.  
The density of molecules at the end of the process is derived for narrow resonance and slow sweep. \end{abstract}

\maketitle

The problem of the molecule production in an experiment where a system of fermionic atoms is tuned from far above to far below the Feshbach resonance 
has recently been studied both experimentally and theoretically. 
First of all, experiments with the systems of ultracold atomic gases close to Feshbach resonance~\cite{Jin2004,Ketterle2004} explored this physics. A number of theoretical 
papers followed~\cite{Altman2005,Barankov2005,Pazy2006,Pokrovsky2006,Stecher2007,Sun:2008qm}.
In the language of the two channel model~\cite{Timmermans1999}, the problem can be formulated in the following straightforward way. Consider the Hamiltonian
\begin{eqnarray}
\label{eq:ham} H &=& \sum_{\p, \sigma} {p^2 \over 2m} ~\hat
a^\dagger_{\p \sigma} \hat a_{\p \sigma} + \sum_p \left(\epsilon_0
+{q^2 \over 4m} \right) \hat b_\q^\dagger \hat b_\q\cr &+&{g \over \sqrt{V}} \sum_{\p,\q}
~ \left( \hat b_\q ~ \hat a^\dagger_{\q/2+\p
\uparrow} \hat a^\dagger_{\q/2-p \downarrow} +
h. c.
\right).
\end{eqnarray}
Here $\hat a^\dagger_{\p \sigma}$, $\hat a_{\p \sigma}$ are creation and annihilation operators of fermions (atoms) of momentum $\p$ and spin $\sigma$, $\hat b^\dagger_\q$, $\hat b_\q$ are creation and annihilation operators of bosons (molecules) with momentum $\q$, $g$ is the coupling, $V$ is the volume, and $\epsilon_0$ is the parameter, called detuning, which controls whether fermions or bosons are energetically favorable. 
The system defined by \rfs{eq:ham} has been extensively studied in the literature, and it is well know that for large positive $\epsilon_0$ this system is a BCS superconductor, while
when $\epsilon_0$ is large and negative, this system describes the Bose condensate of weakly interacting bosonic molecules. 

Now suppose $\epsilon_0$ is a function of time (representing a sweep through the Feshbach resonance), given by
\be \label{eq:sweep}
\epsilon_0=-2 \lambda t,
\ee where $\lambda$ is the rate of its change. Furthermore, suppose initially, at $t \rightarrow -\infty$, the system sits in its ground state which is represented by the Fermi sea
with the Fermi momentum $p_F$
in the absence of the molecules
\be \label{eq:initial} \left< \hat a^\dagger_{\p\sigma} \hat a_{\p\sigma} \right> = \theta(p_F-p), \ \left< \hat b^\dagger_\q \hat b_\q \right> =0, \ t \rightarrow -\infty,\ee
where $\theta(x)$ is the usual step-function, $\theta(x)=1$ if $x>0$ and $\theta(x)=0$ otherwise. One is asked to find the density of molecules at the end of the process, or compute
\be n_b=\frac{1}{V}\sum_q \left< \hat b^\dagger_\q \hat b_\q \right>, \ t \rightarrow \infty.
\ee
By dimensional analysis, the density of the created molecules $n_b$ has to be proportional to the initial density of the fermions, $n_F = p_F^3/(6 \pi^2)$ ($n_F$ is the density of each of the species, spin up or spin down). It can also depend on the two dimensionless parameters in this problem, one being the Landau-Zener parameter \cite{Landau1932,Zener1932,Pokrovsky2006}
\be \label{eq:Gamma} \Gamma = \frac{\pi g^2 n_F}{\lambda} \ee
which controls the rate of the sweep through the Feshbach resonance. The other is the parameter which controls the width of the resonance~\cite{Ho2004,Andreev2004,Gurarie2007a}, 
\be \label{eq:gamma} \gamma = \frac{g^2 m^2}{n_F^{1/3}} \sim \frac{g^2 m^{\frac 3 2}}{ \sqrt{\epsilon_F}}
\ee
Here \be \epsilon_F = \frac{p_F^2}{2m} \ee is the Fermi energy of the initial fermion distribution. 

Thus $n_b=n_F f(\Gamma, \gamma)$, where $f$ is some dimensionless function which needs to be determined. In this paper we analyze the regime where $\gamma \ll 1$, $\ln \left[ 1/(\Gamma \gamma) \right] \ll \Gamma \ll 1/\gamma$, corresponding to a relatively slow sweep of a narrow resonance. Our main result is
\be \label{eq:biganswer}
n_b \simeq n_F \left( 1  - \frac{1}{\Gamma} \ln \frac 1 {\Gamma \gamma} \right).
\ee 

In Ref.~\cite{Pokrovsky2006} the molecular density $n_b$ was calculated  for $\Gamma \ll 1$ perturbatively in powers of $\Gamma$, equvalent to studying a fast sweep. The answer was given by
\be \label{eq:pok} n_b \approx n_F \left( \Gamma - \frac{88}{105} \Gamma^2 + \dots \right)
\ee
(notice that it is independent of $\gamma$ within this order of perturbation theory). 
This should be contrasted with the naive application of the Landau-Zener formula for level crossing \cite{Landau1932,Zener1932} (which requires some generally unjustified replacement of \rfs{eq:ham} by
some sort of a two-level system) which would have given
\be \label{eq:LZnaive} n_b = n_F \left( 1- e^{-\Gamma} \right) \approx n_F \left( \Gamma - \Gamma^2/2+\dots \right).
\ee
Clearly already in the second order of perturbation theory, the Landau-Zener formula breaks down.  
Thus a remaining interesting question would be to compute $n_b$ at large $\Gamma$, the task beyond the reach of the perturbation theory and accomplished in this paper.

%Subsequently, a number of publications offered various competing approaches to this problem. One interesting experimentally relevant regime appears when one replaces 
%\rfs{eq:sweep} by $\epsilon_0=\nu_0>0$ at $t<0$ and $\epsilon_0=\nu_0  -2 \lambda t $ at $t>0$, and this was explored (for relatively fast rate $\lambda$) in Refs~\cite{Altman2005,Barankov2005}. Going back to the problem defined by \rfs{eq:sweep}, one particularly popular approach 

We start our derivation of \rfs{eq:biganswer} with replacing the Hamiltonian \rfs{eq:ham} by a reduced Hamiltonian where only the bosonic zero (momentum) mode is kept \
\be\label{eq:ham1} H = \sum_{\p, \sigma} {p^2 \over 2m} ~\hat
a^\dagger_{\p \sigma} \hat a_{\p \sigma} + \epsilon_0\,
 \hat b^\dagger \hat b +{g \over \sqrt{V}} \sum_{\p}
~ \left( \hat b ~ \hat a^\dagger_{\p
\uparrow} \hat a^\dagger_{-\p \downarrow} +
h. c.
\right).
\ee
This approximation, quote common in the  literature dedicated to this problem \cite{Pazy2006,Sun:2008qm}, is inspired by the fact that one expects a finite fraction of bosons to be Bose condensed at $t \rightarrow \infty$, at least at slow enough sweep $\lambda$. Thus perhaps only the zero mode of the bosons should contribute in a substantial way. 

We are going to argue that going from \rfs{eq:ham} to \rfs{eq:ham1} can only be justified if $\gamma \ll 1$ and $\Gamma \gg 1$. Indeed, it is well known from the studies of the equilibrium problem (where $\epsilon_0$ is time independent) that \rfs{eq:ham1} is a good approximation to \rfs{eq:ham}
if $\gamma \ll 1$, or if the resonance is narrow~\cite{Andreev2004,Gurarie2007a}.  Additionally, at large positive $\epsilon_0$, where the system described by \rfs{eq:ham} is in the BCS regime, \rfs{eq:ham1} is a good approximation regardless of the value of $\gamma$. On the other hand, if $\gamma \gg 1$, then in our problem where  $\epsilon_0$ changes in time from large positive to large negative values, a range of values of $\epsilon_0$ must be traversed, called the unitary regime, where \rfs{eq:ham1} breaks down. 
Thus it is natural to conjecture
 that, if $\gamma \gg 1$, \rfs{eq:ham1} is not a good approximation to \rfs{eq:ham} not only in case of the time independent problem but also for the time dependent sweep \rfs{eq:sweep}.
 
At the same time, if $\gamma \ll 1$, then if $\Gamma \ll 1$ (very fast sweep) a straightforward perturbative calculation for \rfs{eq:ham1} shows that the density of the produced bosons goes as $n_b \approx \Gamma/V$, or in other words, in the thermodynamic limit it vanishes. This is quite different from the analogous result for \rfs{eq:ham},
given by \rfs{eq:pok}. 
This occurs because the molecules in \rfs{eq:ham} formed  under the conditions of a fast sweep are not Bose-condensed. Thus in this regime Eqs.~\rf{eq:ham} and \rf{eq:ham1} describe completely different physics. 

This leaves the parameter range $\gamma \ll 1$, $\Gamma \gg 1$ as the only regime where \rfs{eq:ham} and \rfs{eq:ham1} may be equivalent. And indeed, since under a very slow sweep one expects that most produced molecules are Bose condensed, it is  not unreasonable to assume that this is the case. From now on we adopt this point of view. 

Subsequently, additional common approximation is often made by further neglecting the dispersion of the fermions, to replace \rfs{eq:ham1} with
\be
\label{eq:ham2} H=    \epsilon_0\,
 \hat b^\dagger \hat b + {g \over \sqrt{V}} \sum_{\p}
~ \left( \hat b ~ \hat a^\dagger_{\p
\uparrow} \hat a^\dagger_{-\p \downarrow} +
h. c.
\right).
\ee
We would like to argue that reducing the problem to \rfs{eq:ham2} is problematic. Indeed, it is well known that a time independent version of the problem \rfs{eq:ham2} undergoes a quantum phase transition as a function of $\epsilon_0$~\cite{Dimer2007,Altland2008a}, a feature absent in \rfs{eq:ham} which is widely believed to go through a crossover (termed the BCS-BEC crossover in the literature)~\cite{Leggett1980}.  
Thus this model describes quite a different physics (see however Ref.~\cite{Sun:2008qm} where it is argued that at large $\gamma$, the large $g$ makes it possible to neglect both fermionic and bosonic dispersions and immediately arrive at \rfs{eq:ham2}, in fact in the sector where it does not have a phase transition). 

In this paper, we present the solution to the problem defined by \rfs{eq:ham} with Eqs.~\rf{eq:sweep} and \rf{eq:initial}, with $\gamma \ll 1$  and at a slow sweep $\Gamma \gg 1$, thus justifying replacing \rfs{eq:ham} with \rfs{eq:ham1}.
%This solution combines the techniques 
%of semiclassical equations of motion, first introduced in this context in Refs.~\cite{Barankov2003,Andreev2004,Levitov2004}, with the adiabatic invariants approach of
%Refs.~\cite{Niu2000,Niu2002} which was subsequently exploited to solve the problem defined by \rfs{eq:ham2} in Refs.~\cite{Altland2008,Altland2008a}. 

Next, following the standard approach, we would like to replace the fermion operators by the Anderson pseudospin operators defined by
\be
\hat S^z_p = \oh \left( \hat a^\dagger_{\p \uparrow} \hat a_{p\uparrow} + \hat a^\dagger_{\p \downarrow} \hat a_{p\downarrow} \right), \ \hat S_\p =\hat a_{-\p \downarrow}\hat 
a_{\p \uparrow}, \
\hat S_p^\dagger = \hat a_{\p \uparrow}^\dagger \hat a^\dagger_{-\p \downarrow}.
\ee
We observe that within the model  \rfs{eq:ham1} one can safely replace the quantum operator $\hat b$ by a classical field $b$.  Ref.~\cite{Altland2008a}, which worked with the Hamiltonian \rfs{eq:ham2}, examined this question in a lot of detail. It was established that the classical dynamics was indeed a good approximation to the quantum behavior of \rfs{eq:ham2} with the exception of the early evolution of the system where the number of bosons was close to zero.  A crucial difference, however, between \rfs{eq:ham2} and \rfs{eq:ham1} studied here is the existence of a quantum phase transition as  $\epsilon_0$ in \rfs{eq:ham2} changes. As a result, a pure classical variable $b(t)$, if it is zero initially, remains zero forever within the classical dynamics of \rfs{eq:ham2}, and its quantum behavior must be studied to find $n_b = \left| b \right|^2$ in the infinite future. Fortunately, this does not happen in \rfs{eq:ham1}. 

Thus let us replace the operator $\hat b$ by the classical variable $b$. Once this is done, the spin dynamics   can be described by the classical canonically conjugate variables $n_\p$ and $\phi_\p$ where 
\be S^z_\p = n_\p - \oh, \ S_\p = \sqrt{n_\p (1-n_\p) } \, e^{i \phi_\p}.
\ee
Here $n_\p$ gives the occupation number of a fermion pair at momentum $\p$. 
Taking into account that the Hamiltonian \rfs{eq:ham1} conserves the total number of particles, or that $\sum_\q n_\q +\left| b \right|^2 = n_F V$, we find the effective classical
Hamiltonian equivalent to \rfs{eq:ham1}
\begin{eqnarray} \label{eq:ham3} H &=& \sum_\p \left(2 \epsilon_\p -\epsilon_0 \right) n_p \\ &+&  2 g \sqrt{n_F  -\frac{1}{V} \sum_\q n_\q}  \, \sum_\p \sqrt{n_\p(1-n_\p)} \cos \phi_\p,
\nonumber
\end{eqnarray}
where a convenient notation is introduced, $\epsilon_p=p^2/(2m)$. This Hamiltonian is equivalent to the classical version of \rfs{eq:ham1}, and is supplemented by the
initial conditions, set at $t=-\infty$, stating that $n_\p = \theta(p_F-p)$. 
We could now take advantage of the integrability of \rfs{eq:ham3} 
when $\epsilon_0$ is time independent, discovered recently in Ref.~\cite{Dukelsky2004}. The integrability allows in principle to construct its adiabatic invariants. Then once $\epsilon_0$ depends on time, one could calculate how the adiabatic invariants change with time following \cite{Niu2002}. This technique, although promising, is technically involved and will not be pursued here. Instead, we will follow a related yet somewhat different approach. 

%Its analysis can proceed along the lines first introduced in Ref.~\cite{Niu2002} and exploited for the analysis of \rfs{eq:ham2} in Ref.~\cite{Altland2008a}. 

The structure of the motion described by this Hamiltonian is elucidated by its fixed points, or points where $\d H/ \d n_\p=\d H/ \d \phi_\p=0$. 
These are given by
\be \label{eq:sta}
\phi_\p=\pi, \ n_\p = \oh \left(1-\frac {\xi_\p}{E_\p} \right).
\ee
Here we employ standard notations \be \xi_\p=\epsilon_\p-\mu, \ E_\p = \sqrt{\xi_\p^2 + g^2 \Delta^2}, \ee where $\Delta$ and $\mu$ have to be determined by solving the
two coupled equations
\be \label{eq:pngap}
n_F - \frac 1 V \sum_\q n_\q = \Delta^2, \ \epsilon_0 - 2 \mu = \frac{g^2}{2V} \sum_\p \frac{1}{E_\p}.
\ee
This is the standard BCS-BEC solution of the two channel model, with \rfs{eq:pngap} representing the particle number and the gap equations. In fact, in the limit $\gamma \rightarrow 0$, it is easy to solve these two equations~\cite{Andreev2004,Gurarie2007a}, with the result
\begin{eqnarray} \label{eq:gapsol} \epsilon_0 > 2 \epsilon_F: & \mu=\epsilon_F, &\Delta=0, \cr
2 \epsilon_F>\epsilon_0>0: & \mu = \epsilon_0/2, & \Delta = \sqrt{n_F- \frac{ (2 m \mu )^{3/2}}{6 \pi^2}} \cr
0> \epsilon_0: & \mu = \epsilon_0/2, & \Delta=\sqrt{n_F}.
\end{eqnarray}Notice that if $\gamma>0$, then the solution deviates slightly from \rfs{eq:gapsol}. In particular, at large $\epsilon_0$, $\Delta$ is no longer zero.
Now if $\epsilon_0$ depends on time according to \rfs{eq:sweep} with very small $\lambda$, or $\Gamma \gg 1$, then the mechanical system will adiabatically follow the stationary solution \rfs{eq:sta} with \rfs{eq:gapsol}. This corresponds to the complete conversion of the fermions into bosonic molecules, since the density of molecules at the end of the process will be given by $\Delta^2$ which will be equal, at large negative $\epsilon_0$, to $n_F$. 

However, as the rate $\lambda$ is increased, the number of the produced molecules will decrease. We can describe this in the following way. 
%Let us introduce variables 
%\be \psi^{(1)}_\p= \sqrt{1-n_\p} \, e^{i \phi^{(1)}_\p}, \ \psi^{(2)}_\p = \sqrt{n_\p} \, e^{i \phi^{(2)}_\p},
%\ee where 
%\be \phi_\p = \phi^{(1)}_\p - \phi^{(2)}_\p.
%\ee
%In terms of these variables, the Hamiltonian \rfs{eq:ham3} becomes
%\be H = (\epsilon_p - \frac{\epsilon_0}{2} \right) \left( \left| \psi^{(1)}_\p \right|^2 - \left| \psi^{(2)}_\p \right|^2 \right)
%+ g \left( {\psi^{(1)}_\p}^* \psi^{(2)}_\p +{\psi^{(2)}_\p}^* \psi
Let us compare the Hamiltonian \rfs{eq:ham3} with the standard Hamiltonian of the Landau-Zener two level crossing. That problem can be set up in terms of the two time evolution equations
\be \label{eq:LZ} i \dbyd{\psi_1}{t} = -\lambda t \, \psi_1 + g \, \psi_2, \  i \dbyd{ \psi_2}t = \lambda t \, \psi_2 + g \, \psi_1
\ee
with the initial conditions set at $t \rightarrow -\infty$, $\psi_1=0$ and $\psi_2=1$.
The Hamiltonian for this problem is given by
\be H_{LZ} = \lambda t \left( \left| \psi_2 \right|^2 - \left| \psi_1 \right|^2 \right) + g \left( \psi_1^* \psi_2+\psi_2^* \psi_1 \right).
\ee
Under the mapping  $\psi_1=\sqrt{1-n} \, e^{i \phi_1}$, $\psi_2=\sqrt{n} \, e^{i \phi_2}$, $\phi=\phi_1-\phi_2$ this Hamiltonian can be transformed into
\be
H_{LZ} \label{eq:LZ1} = 2 \lambda t n + 2 g \sqrt{n(1-n)} \cos \phi
\ee 
with the initial condition $n=1$ at $t = -\infty$, 
which possesses a close similarity to \rfs{eq:ham3}. The quantity of interest is $n$ computed at $t=\infty$, known to be equal to
\be \label{eq:LZanswer} \left. n \right|_{t=\infty} =  \exp \left( -\pi g^2/\lambda \right).
\ee

We can then interpret \rfs{eq:ham3}, on the assumption of $\gamma \ll 1$, in the following way. Each $n_\p$ undergoes its own Landau-Zener transition when $\epsilon_0$ is close to $2 \epsilon_\p$ with an effective coupling given by $g \sqrt{n_F-\sum_\q n_\q/V}$. While this process goes through, $n_\p$ for other values of $\p$ do not change appreciably (only one value of $p$ is in `resonance' at a given time), and neither does $\sum_\q n_\q$. Within this interpretation, the final value of $n_\p$ at large positive times, which we denote $n_\p^f$, can be found by comparing \rfs{eq:LZ1} with \rfs{eq:ham3} and using  \rfs{eq:LZanswer}
\be \label{eq:d1} n_\p^f = e^{-\frac{\pi g^2  n_b(p)}{\lambda}},
\ee
Here $n_b(p)$ has a meaning of the boson density at the point in time where $\epsilon_0=2\epsilon_{p}$, and is given by
\be \label{eq:d2} n_b(p) = \frac{1}{V} \sum_{p<q<p_F} (1-n_{\q}^f)+n_{0b},
\ee
where  $n_{0b}$ is the boson density  at $\epsilon_0=2\epsilon_F$. 

Introducing a variable $x=p^3/(6 \pi^2)$, which varies from $0$ to $n_F$, its easy to see that Eqs.~\rf{eq:d1} and \rf{eq:d2} are equivalent to the following equation
\be \dbyd{n_b(x)}{x} + 1 = e^{-\frac{g^2 \pi n_b(x)}{\lambda}},
\ee which leads to the following boson density at the end of the process or when $x=0$
\be \label{eq:nbsol}
n_b=n_b(0) = \frac{\lambda}{\pi g^2} \ln \left[ e^{\frac{\pi g^2 n_F}{\lambda}} \left( e^{\frac{\pi g^2 n_{0b}}{\lambda}} - 1 \right)+1 \right].
\ee
Notice that this argument assumes $\pi g^2 n_{0b}/\lambda \ll 1$. Otherwise, \rfs{eq:nbsol} can predict that $n_b>n_F$ which is contradictory. The origin of this restriction lies in the
fact that at those very slow rates where  this condition is violated, one needs to take into account that the state of the fermions at $\epsilon_0=2\epsilon_F$ is a paired superfluid and not a filled Fermi sea. This leads to a rather complicated problem since \rfs{eq:d1} no holder holds true as initially $n_\p^f$ were not $1$. Considering this goes beyond
the approach developed in this paper.

In the limit $\gamma \rightarrow 0$, $n_{0b}$ is equal to zero. A direct substitution of that into \rfs{eq:nbsol}  leads to $n_b=0$. That means, if we start with zero bosons, we end with zero bosons. This situation is reminiscent of what happens with \rfs{eq:ham2} as discussed at length in Ref.~\cite{Altland2008a}. However, as was pointed out earlier in this paper, our case is crucially different in that, unlike \rfs{eq:ham2} or even \rfs{eq:ham1} in the limit $\gamma=0$, at any finite $\gamma$ our system undergoes a crossover and not a phase transition. To be more precise, during the initial time interval where $\epsilon_0>2 \epsilon_F$, the system will adiabatically follow the solution of \rfs{eq:pngap}. This evolution is very slow at $\gamma \ll 1$, and the deviation from adiabaticity during this time is negligible. 
The value of $\Delta^2$ at $\epsilon_0=2\epsilon_F$ should be used for $n_{0b}$ in \rfs{eq:nbsol}. 
$\Delta^2$ at $\epsilon_0\ge 2 \epsilon_F$ is small but nonzero, and one needs to go beyond the approximation involved in \rfs{eq:gapsol} to find what it is. 
That value can be calculated by carefully solving \rfs{eq:pngap} at finite yet small $\gamma$. Following, for example, the techniques described in Ref.~\cite{Gurarie2007a}
(in particular their equations (6.35) and (6.36)), we find that at $\epsilon_0=2\epsilon_F$, 
\be \Delta^2 = \frac{64 \epsilon_F^2}{e^4 g^2} \exp \left[-\frac{16 \pi^2 \sqrt{\epsilon_F} \kappa} {(2m)^{\frac 3 2} g^2 } \right]=\frac{3 \kappa}{2} n_F,
\ee
where $\kappa=\left(\epsilon_F-\mu\right)/ \epsilon_F$. Solving this equation with logarithmic accuracy at small $\gamma$, we find
\be n_{0b}= \Delta^2 \simeq \frac{3 (2m)^{\frac 3 2} g^2}{32 \pi^2 \sqrt{\epsilon_F}} n_F \sim n_F \gamma.
\ee
We now substitute this value into \rfs{eq:nbsol} under the following conditions $\Gamma \gamma \ll 1$ (which follows from the condition $\pi g^2 n_{0b}/\lambda \ll 1$) and $\Gamma \gg \ln \left[1/(\Gamma \gamma)\right]$, and find
the final answer \rfs{eq:biganswer} which completes the derivation. 

We see that the answer, first of all, implies that it is not as easy to reach the adiabatic behavior by reducing the rate $\lambda$ as it would have been for a Landau Zener level crossing, where the approach to the adiabatic behavior is exponential. Here the number of remaining excitations is linear in $\lambda$, not exponentially suppressed, similarly
to the examples considered in Ref.~\cite{np}. To covert an appreciable fraction of atoms into molecules, the parameter $\Gamma$ must be at least bigger than $\ln(1/\gamma)$, the value which becomes larger (thus the rate smaller) as $\gamma$ decreases. We also notice that \rfs{eq:biganswer} looks similar to some of the expressions found in Ref.~\cite{Pazy2006}, yet the criteria of applicability found in this paper are quite different. 

We also see that the answer \rfs{eq:biganswer} bears some similarity to the behavior of the reduced model \rfs{eq:ham2} studied in Ref.~\cite{Altland2008a}. The similarity appears to have its origin in the fact that \rfs{eq:ham2} undergoes a phase transition as $\epsilon_0$ is varied, while in the limit $\gamma \rightarrow 0$, \rfs{eq:ham1} also has a phase transition, as should be clear from \rfs{eq:gapsol}. Yet at finite $\gamma$, \rfs{eq:ham1} undergoes but a crossover, thus the similarity is somewhat limited. 

Finally,  one is tempted to ask what this implies for the molecule production at large $\gamma \gg 1$, equivalently for broad resonance, the condition under which most of the experiments are done and the majority of theoretical papers dealing with the subject are written. Here one has to admit that the methods developed in this paper are not designed to deal with this problem. We also remark that the well known and experimentally relevant papers \cite{Altman2005,Barankov2005} study a regime quite different from the one studied here (corresponding to $\epsilon_0=\nu_0 - \theta(t)\lambda t$) so that a direct comparison with their results is not possible.

This work was supported by the NSF grant DMR-0449521. The author is grateful to A. Altland and A. Polkovnikov for many useful discussions.

%%%%%%%%% This is the old approach %%%%%%%%%%
%%%%%%%%%%%%%%%%%%%%%%%%%%%%%%%%
%%%%%%%%%%%%%%%%%%%%%%%%%%%%%%%%%

\bibliography{paperLZV}

\end{document}